\documentclass[11pt,letterpaper]{JHEP3}
\usepackage{amsthm}

\newcommand{\nn}{\nonumber} 
\newcommand{\tr}{\mathrm{Tr}} 
\newcommand{\dual}{\star \!} 
\def\dual#1{{\star#1}}

\newcommand{\ie}{{\em i.e., }}
\newcommand{\eg}{{\em e.g., }}

\newcommand{\bc}{\begin{center}}
\newcommand{\ec}{\end{center}}

\newcommand\ZZ{{\mathbb Z}}
\newcommand\RR{{\mathbb R}}

\newfam\scrfam
\batchmode\font\tenscr=rsfs11 \errorstopmode
\ifx\tenscr\nullfont
        \def\scr{\cal}
\else   
        \font\sevenscr=rsfs7
        \font\fivescr=rsfs5
        \skewchar\tenscr='177 \skewchar\sevenscr='177 \skewchar\fivescr='177
        \textfont\scrfam=\tenscr \scriptfont\scrfam=\sevenscr
        \scriptscriptfont\scrfam=\fivescr
        \def\scr{\fam\scrfam}
        \def\cal{\scr}
    \def\mathcal#1{{\scr#1}}
\fi

\title{U-duality covariant membranes}
\author{Viktor Bengtsson, Martin Cederwall, Henric Larsson, Bengt
  E.W. Nilsson\\ Theoretical Physics \\ G\"oteborg University \&
  Chalmers University of Technology \\ S-412 96 G\"oteborg, Sweden \\
E-mail: \email{md9benvi, tfemc, solo, tfebn@fy.chalmers.se}}
\abstract{We outline a formulation of membrane dynamics in $D=8$ which is fully
  covariant under the U-duality group $SL(2,\ZZ)\times SL(3,\ZZ)$, and
  encodes all interactions to fields in eight-dimensional
  supergravity, which is constructed through Kaluza--Klein reduction
  on $T^3$. Among the membrane degrees of freedom is an $SL(2,\RR)$
  doublet of world-volume 2-form potentials, whose quantised electric
  fluxes determine the membrane charges, and
are conjectured to provide an interpretation of the variables
  occurring in the minimal representation of $E_{6(6)}$ 
which appears in the context of automorphic membranes. We solve
  the relevant equations for the action for a restricted class of
  supergravity backgrounds.
Some comments are made on supersymmetry and lower dimensions.
}
\preprint{hep-th/0406223}

\begin{document}

\section{Introduction}

The r\^ole of branes in M-theory is  poorly understood and is in many respects
a very puzzling subject. In string theory the distinction between
fundamental and solitonic branes is clear and provides a conceptually
firm ground  for discussing perturbative as well as non-perturbative issues.
In M-theory this distinction is washed away
due to the inherently non-perturbative nature of the
theory. Nevertheless it can be argued that M2-branes in some sense constitute
microscopic degrees of freedom of M-theory. Matrix theory could be
taken as an argument for this.

By lifting up non-perturbative results from string theory to M-theory
one can get a glimpse of what might be the proper microscopic formulation
of M-theory. Recently the authors of \cite{Pioline:2001jn} suggested a way
forward by trying to derive a well-established exact
non-perturbative string result,
namely the $R^4$-term obtained in a series of papers, see, \eg
\cite{Green:1997tv,Green:1997di,Kiritsis:1997em},
from the M2-brane in M-theory. The functions appearing as partition functions
in this context present a number of interesting features and difficult problems
if one wishes to obtain them from a compactified M2-brane
allowed to wind the compact
target space. Functions of this kind live on the product of two moduli spaces,
one related to the $T^3$ topology of the M2-brane and one connected to the
U-duality group in the lower dimension. One particular property of these
automorphic functions is that their definition requires additional degrees
of freedom on the M2-brane over and above those used in the conventional
formulation of \cite{Bergshoeff:1987cm,Bergshoeff:1988qx}.
The attempt of ref. \cite{Pioline:2001jn} turned out not to give an
entirely correct result, one problem being due to incorrect counting of membrane
instantons when treating them as classical membrane configurations in a
saddle point approximation. The problem was solved in
ref. \cite{Sugino:2001iq} by treating the winding membrane as a
three-dimensional Yang--Mills theory in order to calculate the partition
function. An alternative approach was put forward in
ref. \cite{Pioline:2001jn,Pioline:2004xq}, where it was suggested that a proper
starting point would be a membrane action already manifesting the
U-duality invariance, which then demands the presence of the extra
degrees of freedom mentioned above.

Due to supersymmetry on the M2-brane world-volume these additional degrees of
freedom can not represent local degrees of freedom,
which in fact suggests how they should be introduced.
Field strengths of rank equal to the dimension of the manifold, in
this case the M2 world-volume, have been used to produce exactly modes of
this kind, see, \eg \cite{Aurilia:1980xj}. In string/M theory there is an
extended version of this technique
where all background $n$-form gauge fields $A_{(n)}$
couple to their corresponding $(n-1)$-form gauge fields on the world-volume
$a_{(n-1)}$ through a
universal coupling $f_{(n)}=da_{(n-1)}-A_{(n)}$
\cite{Townsend:1997kr,Cederwall:1997ts,Cederwall:1998ab,Cederwall:1998gg},
sometimes with some additional non-linear terms as we will see below.
Using this method we will in this paper derive an action for the 2-brane in
 8 dimensions which will exhibit manifest U-duality and contain the additional
degrees of freedom that appear in the partition function of Pioline
{\it et al.}
\cite{Pioline:2001jn}. That the additional degrees of freedom are related to
2-form potentials on the M2 world-volume 
was suggested in \cite{Pioline:2004xq}.

In this paper we will use the theory obtained by compactifying 11-dimensional
supergravity on a 3-torus to discuss this problem. We perform the
reduction to 8 dimensions in section 2, where we present
the proper fields needed
to give the Bianchi identities and duality relations in a manifest
$SL(2,\RR)\times SL(3,\RR)$
covariant form. In section 3 we then apply the methods from
\cite{Townsend:1997kr,Cederwall:1997ts,Cederwall:1998ab,Cederwall:1998gg} to
find how the
2-brane couples to
the gauge fields and scalar fields of 8 dimensional supergravity,
including the two
3-form field strengths of special interest here. 
The formalism used gives a set of
field equations and Bianchi identities which are U-duality covariant and
allow for the implementation of the duality relations  needed to define
the theory in terms of the correct number of degrees of freedom.
The discussion is performed throughout
the paper in terms of bosonic fields only.
In order to derive the action of a supersymmetric membrane,
the same formal expressions should be used where all pullbacks to the
brane are taken from a target superspace, as usual.
To identify the new charges
we solve the equations of motion for the world-volume 
2-form potentials in section 4.
In this way
a $(p,q)$ membrane action is obtained which can be compared to the
spherical vector
discussed in \cite{Pioline:2001jn,Kazhdan:2001nx,Pioline:2003bk,Pioline:2004xq}.
Further comments on this connection and
some conclusions are collected in section 5.

\section{Maximal 8-dimensional supergravity}

This section provides a derivation of the 8-dimensional supergravity
obtained by dimensional reduction of $D=11$ supergravity on $T^3$,
with emphasis put on the transformation properties of the fields
under the (continuous version of the)
U-duality group in 8 dimensions, $SL(2,\RR)\times SL(3,\RR)$.
The dynamics of the bosonic sector of $D=11$ supergravity, with vielbein
$\hat{e}_M{}^A$ and 3-form potential $\hat{C}_{MNP}$ (hats are
used on 11-dimensional fields), is given by
the action
\begin{equation}
S = \frac{1}{2\kappa^{2}_{11}}\int d^{11}x \sqrt{|\hat{g}|}\left[
\hat{R} - \frac{1}{2 \cdot 4!} \hat{G}^2\right] +
\frac{1}{2\kappa^{2}_{11}} \int \frac{1}{3!} \hat{G}\wedge\hat{G}\wedge
\hat{C}\ ,
\end{equation}
where $\hat{G}= d\hat{C}$ is the field strength of the 3-form
$\hat{C}$. From now on we set $2\kappa^{2}_{11}=1$.
 Our index conventions are as follows
\begin{equation}
\begin{array}{lll}
& \mathrm{curved} & \mathrm{flat} \\
\mathrm{11-dim} & M ,N ,P, \ldots & A, B, C, \ldots \\
\mathrm{8-dim} & \mu, \nu, \rho, \ldots & a, b, c, \ldots \\
\mathrm{3-dim} & m, n, p, \ldots & i, j, k, \ldots ,
\end{array}
\end{equation}
and our metric is mostly plus. $\hat{G}$ is invariant under the
gauge-transformation
\begin{equation}
\delta \hat{C} = d\hat{\chi}\ ,
\end{equation}
with $\hat{\chi}$ a 2-form. To compactify this theory on $T^3$ we
make the following Kaluza--Klein Ansatz for the vielbein
\begin{equation}
\hat{e}_{M}{}^{A} = \left( \begin{array}{cc}
e_\mu{}^a & -A^{1m}_\mu e_m{}^i \\
0 & e_m{}^i
\end{array} \right),\quad
\hat{e}_{A}{}^M = \left( \begin{array}{cc}
e_a{}^\mu & e_a{}^\mu A^{1m}_\mu \\
0 & e_i{}^m
\end{array} \right),
\label{eq:defofehat}
\end{equation}
\begin{equation}
dx^{M} = (dx^\mu, dx^m)\ ,
\end{equation}
where all fields are allowed to depend on $x^\mu$ only.
This gives
\begin{equation}
\hat{e}^{A} = dx^M\hat{e}_M{}^A
= (dx^\mu e_\mu{}^a,  dx^m e_m{}^i -dx^\mu A^{1m}_\mu e_m{}^i ) = (e^a,
  e^i-A^{1i}) = (\hat{e}^a, \hat{e}^i)\ .
\end{equation}
The index $1$ on $A^{1m}_{\mu}$ is needed to separate this 
vector from the one that will arise from the compactified 3-form.

Using the above information we find, after some calculations, that
the 11-dimensional Einstein term gives the following contribution
to the 8-dimensional action:

\begin{eqnarray}\label{r118}
\int d^{11}x \sqrt{|\hat{g}|}\hat{R} &=& \int d^{8}x
\sqrt{|g|}e^{-\varphi} \left[ R -
\frac{1}{4}G_{mn}F^{1m}_{ab}F^{1n\,ab} \right.\nn \\
&-& \left. 2D_{a} (e_i{}^{m}
\partial^{a}
e_m{}^{i}) -  e^{n(i} (\partial_{a} e_n{}^{j)})e_i{}^{m}(\partial^{a} e_{mj})
- e_i{}^{m}(\partial^{a}
e_m{}^{i}) e_j{}^{n}(\partial_{a} e_n{}^{j}) \right],  \nn
\end{eqnarray}
where we have used the definitions
\begin{equation}
\sqrt{\det{G_{mn}}} = e^{-\varphi}\ ,\quad
F^{1m}_{ab}=dA^{1m}_{ab}\ .
\end{equation}
Note also that $G_{mn}$ is the metric on $T^3$ constructed from the
dreibein ${e_m}^i$
and that the derivatives have been written with flat indices
using the 8-dimensional vielbein.
We have put the parametric volume  of the
internal torus to one, $\int d^3y=1$.
The third term in the action can be integrated by parts
\begin{eqnarray}\label{partint}
-2 \int d^{8}x \sqrt{|g|} e^{-\varphi} D_{a} (e_i{}^{m}
\partial^{a} e_m{}^{i}) &=& 2 \int d^{8}x \sqrt{|g|}
e^{-\varphi} (\partial_{a} \varphi)(\partial^{a} \varphi)\ ,
\end{eqnarray}
where we have used the fact that
\begin{equation}
\tr (e^{-1} \partial e) = -\partial \varphi\ .
\end{equation}
Next, inserting (\ref{partint}) into (\ref{r118}) gives
\begin{eqnarray}
\int d^{11}x \sqrt{|\hat{g}|}\hat{R} &=& \int d^{8}x \sqrt{|g|}e^{-\varphi}
\left[ R - \frac{1}{4}G_{mn}F^{1m}_{ac}F^{1n\,ac} + (\partial_{a}
\varphi)(\partial^{a}
\varphi) \right.\nn \\
&-& \left.  e^{n(i} (\partial_{a} e_n{}^{j)})e_j{}^{m}
(\partial^{a} e_{mi}) \right].
\end{eqnarray}
One now realises that the last term is most conveniently rewritten as
\begin{eqnarray}
&-& \frac{1}{2} e^{ni} (\partial_{a} e_n{}^{j})e_j{}^{m}(\partial^{a} e_{mi})
- \frac{1}{2} e^{ni} (\partial_{a} e_n{}^{j})e_i{}^{m}(\partial^{a}
e_{mj})
\nn \\
&=& -\frac{1}{4}
 (G^{mn} \partial_{a} G_{np})(G^{pq} \partial^{a} G_{qm})
= -\frac{1}{4} \tr (G^{-1} \partial G)^{2}\ ,
\end{eqnarray}
giving the action
\begin{eqnarray}
\int d^{11}x \sqrt{|\hat{g}|} \hat{R} = \int d^{8}x \sqrt{|g|}
e^{-\varphi}
\Bigl[ R + (\partial_{a} \varphi)(\partial^{a} \varphi)\Bigr. &-&
\frac{1}{4}G_{mn}F^{1m}_{ab}F^{1n\,ab}\nn \\
 &-& \Bigl.\frac{1}{4} \tr (G^{-1} \partial G)^{2} \Bigr]\ .
\end{eqnarray}
With the calculation of the Ricci scalar out of the way we would now
like to rescale the result to obtain an action in the Einstein
frame. We perform the change of variables
\begin{equation}
g_{\mu \nu} = e^{\varphi /3} g_{\mu \nu}^{\rm E}\ ,\quad
{\cal M}_{mn} = \frac{G_{mn}}{(\det{G})^{1/3}}\ ,
\end{equation}
where the three by three matrix ${\cal M}_{mn}$ provides
a parametrisation of the
 $SL(3,\mathbb{R})/SO(3,\mathbb{R})$ coset space.
This gives the final form of the compactified  Einstein--Hilbert term
\begin{eqnarray}
\int d^{11}x \sqrt{|\hat{g}|} \hat{R} = \int d^{8}x
\sqrt{|g^{\mathrm{E}}|}
\Bigl[ R_{\mathrm{E}} - \frac{1}{2} (\partial \varphi)^{2}&-&\frac{1}{4}
e^{-\varphi} {\cal M}_{mn} F^{1m} F^{1n} \nn\\
&-&\frac{1}{4} \tr (\partial
{\cal M} {\cal M}^{-1})^{2} \Bigr]\ .
\end{eqnarray}

Next, we reduce the 4-form field strength. This is most easily done
by expressing the forms in the flat 8-dimensional basis and the
``real'' (according to our Kaluza--Klein Ansatz) curved 3-dimensional basis
\begin{equation}
\hat{e}^{m} = dx^{m} + A^{1m}\ , \quad A^{1m} = dx^{\mu}A^{1m}_{\mu}
\ ,
\end{equation}
satisfying
$\hat e^{m} e^{i}_{m} = \hat e^{i}$,
as well as
$d \hat{e}^{m} = d A^{1m} = F^{1m}$.
This basis is invariant under transformations $\delta x^{m} =
-\lambda^{m}$
of the internal torus as is seen from
\begin{eqnarray}
\delta \hat{e}^{m} &=& \delta(dx^{m} + dx^{\mu} A^{1m}_{\mu} )
= d \delta x^{m}
+ dx^{\mu}(\delta A^{1m}_{\mu})
\nn \\
&=& - d \lambda^{m} + dx^{\mu}(\partial_{\mu} \lambda^{m})  = - d
\lambda^{m} + d \lambda^{m} = 0\ ,
\end{eqnarray}
where we have used
\begin{eqnarray}
\delta A^{1m}_{\mu} = \partial_{\mu} \lambda^{m}\ .
\end{eqnarray}
This is the usual way reparametrisations generate  gauge transformation
in a Kaluza--Klein reduction.
When expanding the 11-dimensional 3-form into lower-dimensional
components (8-dimensional in our case)
it is convenient, and completely standard, to perform field
redefinitions of the various fields to get them
inert under this gauge invariance. Using  our basis defined above, and the
fact that $\hat{C}$ is
manifestly invariant, we conclude that the fields appearing in the
expansion (\ref{hatC})
below, do in fact not transform either. Note that the gauge fields
$A^{1}$ in $\hat{e}^{m}$
will not appear explicitly anywhere since when integrating
out the three compact directions from the action only terms with
three $dx^{m}$'s
will be non-zero. Therefore  $\hat{e}^{m}$ appears as $dx^{m}$ from the point
of view of the action.

Let us begin by reducing the $D=11$ relation
\begin{eqnarray}
\delta \hat{C} = d \hat{\chi}\ .
\end{eqnarray}
As explained above, we expand the 11-dimensional 3-form into $\lambda^{m}$
invariant fields as follows:
\begin{eqnarray}\label{hatC}
\hat{C} = C' + B'_{m} \wedge \hat{e}^{m} + \frac{1}{2!} A^{2p} \wedge
\epsilon_{mnp} \hat{e}^{m} \wedge \hat{e}^{n} + \frac{1}{3!} a
\epsilon_{mnp} \hat{e}^{m} \wedge \hat{e}^{n} \wedge \hat{e}^{p}\ ,
\end{eqnarray}
where $\epsilon_{mnp}$ is defined to be constant.
Repeating this  for the gauge parameter, using the superspace
convention of exterior
derivatives acting from the right, we find
\begin{eqnarray}
d \hat{\chi} &=& d(\chi' - \chi'_{m} \wedge \hat{e}^{m} + \frac{1}{2!}
\epsilon_{mnp}\chi'^{2p} \hat{e}^{m} \wedge \hat{e}^{n}) \nn \\
&=& \underbrace{d\chi' - \chi'_{m} \wedge F^{1m}}_{\delta C'} +
(\underbrace{d\chi'_{m} +
\epsilon_{mnp}\chi'^{2p} F^{1n}}_{\delta B'_{m}}) \wedge
\hat{e}^{m} \\
&+& \underbrace{d\chi'^{2p}}_{\delta A'^{2p}}\wedge
\epsilon_{mnp}\frac{1}{2!}\hat{e}^{m} \wedge \hat{e}^{n}\ ,\nn
\end{eqnarray}
where we have also indicated which field transformations the terms
are connected to.
Moreover, one immediately finds the field strengths that are invariant
under the above
$\lambda^{m}$ transformations as follows:
\begin{eqnarray}
\hat{G} = d \hat{C} &=& d(C' + B'_{m} \wedge \hat{e}^{m} +
\frac{1}{2!}A^{2p} \wedge \epsilon_{mnp} \hat{e}^{m} \wedge \hat{e}^{n} \nn
+\frac{1}{3!} a \epsilon_{mnp} \hat{e}^{m} \wedge \hat{e}^{n}
\wedge \hat{e}^{p}) \\
&=& \underbrace{dC' + B'_{m} \wedge F^{1m}}_{G} + \underbrace{(-d B'_{m}
+ A^{2p} \epsilon_{mnp} \wedge F^{1n})}_{-H_{m}} \wedge \hat{e}^{m}  \\
&+&\frac{1}{2!}\underbrace{(dA^{2p}
+a F^{1p})}_{F'^{2p}} \epsilon_{mnp} \wedge \hat{e}^{m} \wedge \hat{e}^{n}
- \frac{1}{3!} (da) \epsilon_{mnp} \wedge \hat{e}^{m} \wedge
\hat{e}^{n} \wedge \hat{e}^{p}\ .\nn
\end{eqnarray}
Hence, our 8-dimensional field strengths become:
\begin{eqnarray}
G&=&dC'+B'_{m} \wedge F^{1m}\ ,\nn\\
H_{m}&=&d B'_{m} -  \epsilon_{mnp}F^{1n}\wedge A^{2p}\ ,\label{field}\\
F'^{2m}&=&F^{2m}+aF^{1m}=dA^{2m}+aF^{1m}\ ,\nn
\end{eqnarray}
satisfying the following non-trivial Bianchi identities
\begin{equation}
dG=H_{m} F^{1m}\ ,\quad dH_{m}=- \epsilon_{mnp} F^{1n}\wedge
F^{2p}\ ,\quad dF'^{2m}=da\wedge F^{1m}\ .
\end{equation}

Next, we reduce the $\hat{G}^{2}$ term in the 11-dimensional
action. Using (\ref{field}) and rewriting the $\hat{G}^{2}$ term
in terms of forms
\begin{equation}
S_{\hat{G}^{2}}=-\int\frac{1}{2}\hat{G}(\ast_{11}\hat{G})\ ,
\end{equation}
we obtain, after a short calculation, the following contribution
to the 8-dimensional action:
\begin{eqnarray}
S_{\hat{G}^{2}}=-\int d^{8}x\sqrt{|g_{\mathrm{E}}|}
\Bigl[\frac{1}{48}e^{-\varphi}G^{2}&+&\frac{1}{12}H_{m}H_{n}{\cal M}^{mn}
\nn\\
&+&\frac{1}{4} e^{\varphi} {\cal M}_{mn} F'^{2m} F'^{2n} +
\frac{1}{2}e^{2\varphi}(\partial a)^{2}\Bigr]\ , \label{eq:sugrapart1}
\end{eqnarray}
where $G^2=G_{\mu\nu\rho\sigma}G^{\mu\nu\rho\sigma}$.
In the final result we have also switched to the Einstein metric.
Note also that when reducing, the $\ast_{11}$ splits nicely into
$(\ast_{8})(\ast_{3})$. The next step is to write the
Einstein term together with the $\hat{G}^{2}$ term in a more
$SL(2,\mathbb{R})\times SL(3,\mathbb{R})$ covariant way. This leads to
\begin{eqnarray}\label{sl2sl31}
\int d^{11}x \sqrt{|\hat{g}|} (\hat{R}&-&
\frac{1}{48}\hat{G}^{2})\nn\\
&=& \int d^{8}x \sqrt{|g_{\mathrm{E}}|} [
R_{\mathrm{E}} -\frac{1}{4} \tr (\partial {\cal W}
{\cal W}^{-1})^{2} - \frac{1}{4} \tr (\partial
{\cal M} {\cal M}^{-1})^{2} \nn \\
&-&\frac{1}{4}{\cal M}_{mn} F^{rm} F^{sn}{\cal
W}_{rs}-\frac{1}{12}H_{m}H_{n}{\cal M}^{mn} -
\frac{1}{48}e^{-\varphi}G^{2}]\ ,
\end{eqnarray}
where $r,s=1,2,$ and ${\cal W}$ is the following metric parametrising the
$SL(2,\mathbb{R})/SO(2)$ coset:
\begin{equation}\label{W}
{\cal W}=\frac{1}{{\rm Im}(\tau)}\pmatrix{|\tau|^{2} & {\rm
Re}(\tau)\cr {\rm Re}(\tau) & 1\cr }
=e^\varphi\pmatrix{a^2+e^{-2\varphi}&0\cr a&1\cr}\ ,
\end{equation}
with $\tau=a + ie^{-\varphi}$.
Excluding the $G^{2}$ term, it is easy to see that (\ref{sl2sl31})
is invariant under the following $SL(2,\mathbb{R})$ and
$SL(3,\mathbb{R})$
transformations:
\begin{equation}\label{sl22}
{\cal W} \rightarrow \Lambda{\cal W}\Lambda^{\rm T}\ ,\quad F^{m}
\rightarrow (\Lambda^{\rm T})^{-1}F^{m}\ , \quad \Lambda\in SL(2)\
,
\end{equation}
\begin{equation}\label{sl33}
{\cal M} \rightarrow R{\cal M}R^{\rm T}\ ,\quad H_{m} \rightarrow
R_{m}{}^{n} H_{n}\ , \quad F^{m} \rightarrow
F^{n}(R^{-1})_n{}^{m}, \quad R\in SL(3)\ .
\end{equation}
The metric in the Einstein frame is invariant under both
transformations while the  4-form $G$
requires a separate discussion which will be presented after treating
the Chern--Simons term.

In writing down the Chern--Simons term we use that the terms
resulting from the product must not have more than three 3-beins
(or equivalently after integrating out the compact directions a
form of degree eight must remain), all other terms are trivially
zero. This gives
\begin{eqnarray}
\int \frac{1}{6} \hat{G} \wedge \hat{G} \wedge \hat{C} &=& \int d^8x
\frac{1}{6^{3} \cdot 2^{4}} \left[ GGa + 8 G H_{m} A^{2m} + 12
G F'^{2m} B'_{m} \right. \nn \\
&+& \left. 8 G (da) C' - 8 H_{m} H_{n}
B'_{p} \epsilon^{mnp} + 16 H_{m} F'^{2m} C' \right]\ ,
\label{eq:sugrapart2}
\end{eqnarray}
where all space-time indices have been suppressed 
(the eight indices on each term are understood as being contracted
with an epsilon-tensor).
In conclusion the complete action for $D=11$ supergravity reduced
to 8 dimensions is given by (\ref{sl2sl31}) and (\ref{eq:sugrapart2}).

Up to this point we have followed closely the presentation in
\cite{Alonso-Alberca:2000gh}. However, the discussion below will
differ somewhat adding a couple of clarifying points to previous work.

Before we continue and reduce the 7-form we make the following
observation: Looking at how the field strength $H_{m}$ is defined above,
we see that it is not written in an $SL(2,\mathbb{R})$ invariant way. This is
fixed by redefining the potential $B'_{m}$ as follows:
\begin{equation}
B'_{m}=B_{m}-\frac{1}{2}\epsilon_{mnp}A^{1n}\wedge A^{2p}\ ,
\end{equation}
which implies that
\begin{equation}
H_{m}=dB_{m}-\frac{1}{2}\epsilon_{mnp}(F^{1n}\wedge A^{2p} -
F^{2n}\wedge
A^{1p})=dB_{m}-\frac{1}{2}\epsilon_{mnp}\epsilon_{rs}F^{rn}\wedge
A^{sp}\ ,
\end{equation}
where $r,s=1,2.$ It is also convenient to redefine $C'$ as
follows:
\begin{equation}
C'=C-\frac{1}{3}A^{1m}\wedge B_{m} +
\frac{1}{6}\epsilon_{mnp}A^{1m}\wedge A^{2n}\wedge A^{1p}\ ,
\end{equation}
which implies that
\begin{equation}
G=dC+\frac{2}{3}B_{m}\wedge F^{1m} - \frac{1}{3}A^{1m}\wedge
H_{m}\ .
\end{equation}
Note that the Bianchi identities are of course not changed by these
redefinitions. It is clear, however, that the new $B$ and $C$
fields do transform
under reparametrisations along the internal 3-torus which is unavoidable
consequence of imposing manifest $SL(2)$ covariance.

Next we are going to show how the 11-dimensional 7-form
reduces to, among other things, a 4-form, denoted as $G'$, 
which is dual to the 4-form
$G$. The 11-dimensional 4- and 7-forms can be expanded as follows:
\begin{eqnarray}\label{47}
\hat{G} = G - H_{m} \wedge \hat{e}^{m} + \frac{1}{2!} F'^{2p} \wedge
\epsilon_{mnp} \hat{e}^{m} \wedge \hat{e}^{n} - \frac{1}{3!} da \wedge
\epsilon_{mnp} \hat{e}^{m} \wedge \hat{e}^{n} \wedge \hat{e}^{p}\
, \nn \\
\hat{G}_{7} = G_{7} - G_{6m} \wedge  \hat{e}^{m} + \frac{1}{2!} G_{5mn} \wedge
\hat{e}^{m} \wedge \hat{e}^{n} - \frac{1}{3!} G' \wedge \epsilon_{mnp}
\hat{e}^{m} \wedge \hat{e}^{n} \wedge \hat{e}^{p}\ .
\end{eqnarray}
From $\hat{G}_{7}$ we are only interested in the 4-form $G'$ (since
higher tensors do not couple to membranes),
which is the field strength we need to form an $SL(2)$ doublet
together with $G$.

We begin by deriving the Bianchi identity for $G'$, using the
Bianchi identity for $\hat{G}_{7}$
\begin{equation}
d\hat{G}_{7}=\frac{1}{2}\hat{G}\wedge \hat{G}\ .
\end{equation}
Next, from the above Bianchi identity and (\ref{47}), we obtain
the following Bianchi identity for $G'$:
\begin{equation}
dG'=G\wedge da + H_{m}\wedge F'^{2m}=G\wedge da + H_{m}\wedge
F^{2m} + H_{m}\wedge F^{1m}a\ .
\end{equation}
We also find that the duality relation
$\hat{G}_{7}=\ast_{11}\hat{G}$ gives the following duality
relation between the two 4-forms in eight dimensions:

\begin{equation}
G'=-e^{-\varphi}(\ast_{8}G)\ .
\end{equation}
One slight disadvantage with the way we have defined $G'$ is that
it is difficult to write the two 4-forms as an $SL(2,\mathbb{R})$
doublet. To rectify this we define a new 4-form $\tilde{G}$ as
\begin{equation}\label{G44}
\tilde{G}=G'-aG=-e^{-\varphi}(\ast_{8}G)-aG\ .
\end{equation}
This implies that the Bianchi identity for $\tilde{G}$ is given by
\begin{equation}
d\tilde{G}=H_{m}\wedge F^{2m}\ .
\end{equation}
Hence, we can write the Bianchi identities for the two 4-forms
$\tilde{G}$ and $G$ in the following $SL(2,\mathbb{R})$ covariant form:
\begin{equation}
dG^{r}=H_{m}\wedge F^{rm}\ ,
\end{equation}
where we have defined $G^{1}=G$, $G^{2}=\tilde{G}$, $r=1,2,$
$F^{2m}=dA^{2m}$ and
\begin{equation}
G^{r}=dC^{r}+\frac{2}{3}B_{m}\wedge F^{rm} -
\frac{1}{3}A^{rm}\wedge H_{m}\ .
\end{equation}

Moreover, using the $SL(2,\mathbb{R})$ covariant notation for the
two 4-forms implies that the duality relation (\ref{G44}) can
be conveniently rewritten in the following way
\begin{equation}
\ast_{8}G^{r}=-\epsilon^{rs}G_{s}=-\epsilon^{rs}{\cal
W}_{st}G^{t}\ ,
\end{equation}
where $\epsilon^{12}=1$ and ${\cal W}$ is the symmetric
$SL(2,\mathbb{R})/SO(2)$ matrix given above in (\ref{W}).

Furthermore, in this $SL(2,\mathbb{R}) \times SL(3,\mathbb{R})$
covariant notation the variation of the potentials is given by:
\begin{eqnarray}\label{gt1}
\delta A^{rm}&=&d\chi^{rm}\ ,\quad  \delta
B_{m}=d\chi_{m}-\frac{1}{2}\epsilon_{mnp}\epsilon_{rs}A^{rn}\wedge
d\chi^{sp}\ ,\nonumber\\
\delta C^{r}&=&d\chi^{r}-\frac{2}{3}A^{rm}\wedge
d\chi_{m}+\frac{1}{3}B_{m}\wedge
d\chi^{rm}+\frac{1}{6}\epsilon_{mnp}\epsilon_{st}A^{rm}\wedge
A^{sn}\wedge d\chi^{tp}\ ,
\end{eqnarray}
where $\chi^{rm}$ is a 0-form, $\chi^{m}$ a 1-form and $\chi^{r}$
a 2-form. The field strengths $G^{r}$, $H_{m}$ and $F^{rm}$, are
gauge invariant under the gauge transformation (\ref{gt1}).

The $SL(2,\mathbb{R})$-doublet we have derived above can be encoded in
a slightly more elegant manner by using complex
$SL(2,\mathbb{R})$-zweibeins. These are defined by a complexification
of the real zweibeins which are in turn extracted from the
$SL(2,\mathbb{R})$-metric, ${\cal W}$, according to
\begin{eqnarray}
{\cal W}_{rs} = \nu_r{}^{I} \nu_s{}^{J} \delta_{IJ}\ ,
\end{eqnarray}
giving us (modulo $SO(2)$ transformations, \ie in a certain $SO(2)$ gauge)
\begin{eqnarray}
\nu_1{}^{1} = a e^{\varphi/2},\quad \nu_1{}^{2}
= e^{-\varphi/2}\ ,\quad  \nu_2{}^{1} = e^{\varphi/2}\ , 
\quad \nu_2{}^{2} = 0\ .
\end{eqnarray}
Complexifying these,
${\cal U}_{r} = \nu_{r}{}^{1} + i\nu_{r}{}^{2}$,
then gives
\begin{eqnarray}
{\cal U}_{1} &=& a e^{\varphi/2} + ie^{-\varphi/2}=e^{\varphi/2}\tau\ , \\
{\cal U}_{2} &=& e^{\varphi/2}\ .
\end{eqnarray}
The $SL(2,\mathbb{R})$-indices can then be absorbed using these
zweibeins giving us
\begin{eqnarray}
{\cal G} &=& {\cal U}_{r}G^{r} =  e^{\varphi/2}(aG^{1} + G^{2})
+ ie^{-\varphi/2}G^{1}\ , \\
{\cal F}^{m} &=& {\cal U}_{r}F^{rm} =  e^{\varphi/2}(aF^{1} +
F^{2}) + ie^{-\varphi/2}F^{1}\ .
\end{eqnarray}
Similarly, $SL(3,\RR)$-invariant fields are obtained by contracting
with dreibeins ${\cal V}_m{}^i$ fulfilling
${\cal V}{\cal V}^T={\cal M}$.
Converting from $SL$ to $SO$ hides the ``metric'' information in the
action and the $F^2$ term, for example, takes the form ${\cal
  F}^i\bar{\cal F}^i$.
Bianchi identities for $SL$-invariant field strengths include the
left-invariant Maurer--Cartan forms for ${\cal U}$ and ${\cal V}$, and
the kinetic terms of scalars may be written as the square of the part
of the Maurer--Cartan form outside the gauged $so$ algebra.
This $SL$-invariant notation, in addition to a certain amount of
elegance, becomes necessary when we want to consider fermionic fields and
supersymmetry. The $U(1)$ acting on the complex zweibeins is also the
one that rotates the complex spinors in $D=8$.

To summarise this section, we here collect some useful formulas, namely
the manifest U-duality covariant Bianchi identities for the 2-, 3-, and 4-form
field strengths, with $\epsilon_{12}=+1$,
\begin{equation}
dG^r=H_{m} \wedge F^{rm}\ ,\quad
dH_{m}=- \frac{1}{2} \epsilon_{rs}\epsilon_{mnp} F^{rn}\wedge
F^{sp}\ ,\quad dF^{rm}=0\ ,
\end{equation}

and the 8-dimensional duality relation for the 4-form,
\begin{equation}
\ast_{8}G^{r}=-\epsilon^{rs}G_{s}=-\epsilon^{rs}{\cal
W}_{st}G^{t}\ .
\end{equation}

\section{U-duality covariant membrane dynamics }

 The aim of this section is to write down an action for a membrane
 that couples to all the fields in the supergravity background derived
 in the previous chapter. This is done via world-volume field strengths
 roughly of the form ``$f = da - A$'', where $a$ is a world-volume field
 and $A$ the pullback of a background field. Any background tensor
 field (of low enough rank to couple to a membrane) has its
 world-volume counterpart. This procedure was originally invented for
a somewhat different purpose \cite{Bergshoeff:1992gq}, but it soon became
 evident that it has several interesting features: it encodes the
 background coupling, and thereby the way branes may end on each other,
 in a covariant manner, and, in cases where branes themselves come in
 multiplets of a symmetry group of the supergravity, a single action
 encodes all charge sectors. The formalism was developed and
 generalised in refs.
\cite{Townsend:1997kr,Cederwall:1997ts,Cederwall:1998ab,Cederwall:1998gg}.
The last property is what is interesting
 to us here; it will allow us to formulate membrane dynamics in a way
 that accounts for the two types of membranes occurring in 8
 dimensions, namely the direct reduction of the M2-brane and the
 winding M5-brane, carrying electric charges with respect to the two
 3-form gauge fields in the supergravity theory
(or, considering self-duality, electric
 and magnetic charge, respectively, under one of them). These charges
 are identified in terms of electric fluxes of an $SL(2)$ doublet of
 2-form potentials on the membrane. This identification allows for a
 direct physical interpretation of the variables occurring in
 non-linear realisation of the automorphic membrane group $E_{6(6)}$
\cite{Pioline:2001jn,Kazhdan:2001nx,Pioline:2004xq,Gunaydin:2000xr}.

A formulation of brane dynamics where every background field has a
world-volume counterpart will automatically be covariant with respect
to the global symmetries of the background (ungauged) supergravity to which the
brane couples, \ie under the U-duality group. The principal form of
the background coupling, ``$f = da - A$'' (which in the case of
modified Bianchi identities in the background will be suitably
modified, see below), and of the background gauge
transformations $\delta A=d\Lambda$, 
which are accompanied by a shift in the world-volume
potential, $\delta a=\Lambda$, is directly related to the
general nature of world-volume fields being Goldstone modes
corresponding to background ``gauge symmetries'', that become global
symmetries for parameters taking non-zero values on the brane
\cite{Adawi:1998ta}\cite{Cederwall:1998tr}.

As will be exemplified below, this procedure generically leads to a
situation where the number of world-volume fields is larger than the
number of physical Goldstone modes. Constraints must be placed on the
fields, typically in the form of some self-duality condition. By
making a general enough Ansatz for the action and demanding that the
constraints are consistent with the way the background fields occur in
Bianchi identities and in equations of motion determines the
action. In the present case, as will be clear below, we have not
succeeded in solving this problem for general backgrounds, due to a
certain non-linearity of the constraints. This will
not affect our general conclusions, but means that we have not
strictly speaking achieved a formulation that is covariant under the
full U-duality group for general $D=8$ supergravity backgrounds.
The technical details are explained later in
this section.

The governing principle in writing down the field strengths is of
course gauge invariance. Having modified Bianchi identities in the
background will demand that we add non-trivial terms to $da - A$ to
account for the way that the background potentials transform. A
convenient form of the modifying terms is ``world-volume potential
$\times$ background field strength''. A short calculation reveals
\begin{eqnarray}
w^{rm} &=& d \phi^{rm} - A^{rm}\ , \nn \\
f_{m} &=& d a_{m} - B_{m} + \frac{1}{2} \varepsilon_{mnp}
\varepsilon_{rs} \phi^{rn} F^{sp}\ , \\
h^{r} &=& db^{r} - C^{r} - \frac{1}{3} \phi^{rm} H_{m} - \frac{2}{3}
a_{m} \wedge F^{rm} \\
&+& \alpha w^{rm} \wedge f_{m} + \beta
\varepsilon_{mnp} \varepsilon_{st} {\cal W}^{rs} {\cal W}_{uv}
w^{tm}\wedge w^{un}\wedge w^{vp}\ , \nn
\end{eqnarray}
These field strengths are invariant under the target space 
gauge transformations in (\ref{gt1}) together with
\begin{eqnarray}
\delta \phi^{rm} &=& \chi^{rm}\ , \nn \\
\delta a_{m} &=& \chi_{m} - \frac{1}{2} \varepsilon_{mnp}
\varepsilon_{rs} A^{rn} \chi^{sp}\ , \\
\delta b^{r} &=& \chi^{r} - \frac{2}{3} A^{rm} \wedge \chi_{m}
+ \frac{1}{3} B_{m} \chi^{rm} + \frac{1}{6} \varepsilon_{mnp}
\varepsilon_{st} A^{rm} \wedge A^{sn} \chi^{tp}\ . \nn
\end{eqnarray}
Below, we will make use of the following Bianchi identities
\begin{eqnarray}
dw^{rm} &=& - F^{rm}\ , \label{eq:biwv1} \\
df_{m} &=& - H_{m} + \frac{1}{2} \varepsilon_{mnp} \varepsilon_{rs}
F^{rn} \wedge w^{sp}\ . \label{eq:biwv2}
\end{eqnarray}
The last two terms in the definition of $h^r$, containing the
parameters $\alpha$ and $\beta$, yet to be determined, are obviously gauge
invariant on their own accord. Their inclusion is explained
below. Note that the $\beta$-term, which in the real formalism we use
looks a bit complicated, in complex formalism equals the simple
expression
$i\beta \varepsilon_{mnp}w^{m}\wedge w^{n}\wedge \bar w^{p}$.

Counting the number of degrees of freedom gives a too high
number. Apart from the five transverse scalars, a
supersymmetric membrane should only have three additional bosonic degrees of
freedom, which can be taken as for example the triplet of internal
scalars $\phi^{1m}$. The doublet of 2-forms do not contain any local
degrees of freedom, but the remaining scalars and the vectors should
be related to the physical scalars by some relation. This relation is
a duality relation, as will be explained in a little while. The
``action'' we will write down initially does not have this duality
relation as an equation of motion, but is consistent with it. This
situation has been encountered earlier in refs.
\cite{Cederwall:1998ab,Cederwall:1998gg} .

The actions for various branes in the type of formulation we use is
always of the form ``$S\sim\int\lambda(1+\Phi+h^2)$'' where $\Phi$ is
some polynomial function of world-volume field strengths (except the
maximal ones). They are quadratic in the max-forms $h$.  
The variable $\lambda$ is a scalar Lagrange
multiplier. Note that
there is no Wess--Zumino term; that coupling is instead reproduced by
the $h^2$ term once the equation of motion for $\lambda$ is used to
eliminate $h$.
We thus write
\begin{eqnarray}
S = \int d^{3} \xi \sqrt{-g} \lambda \left[1 + \Phi(w,f) -
  {\dual h^{r}}{\dual h^{s}} {\cal W}_{rs}\right]\ , \label{eq:actanz}
\end{eqnarray}
where $\Phi$ is a, yet unknown, function we aim to obtain. The method
for determining its exact form is to consider consistency of the
background couplings encoded in the field strengths with some duality
relation. This is where the last two terms in the definition of $h^r$
enter; they are needed for consistency of the (any) duality. We also
note that we could equivalently leave them out and instead introduce
terms linear in $h$ in the action.

We begin by deriving the equations of motion from the above action.
They will, due to the fact that we do not know the form of $\Phi$, be
implicit. By taking into account how the potentials enter the various
field strengths, we get
\begin{eqnarray}
\phi^{rm} &:& d \left[ \lambda \dual j_{rm} - 2\alpha  \lambda f_{m}
  \wedge \dual h^{s} {\cal W}_{sr} - 2 \beta  \lambda
  \varepsilon_{mnp} \left( \varepsilon_{sr} {\cal W}^{vs} {\cal
  W}_{ut} + 2 \varepsilon_{st} {\cal W}^{vs} \mathcal{W}_{ur}
\right) w^{un} w^{tp} \dual h^{s'} \mathcal{W}_{s'v} \right] \nn \\
&=& \lambda \left[- \frac{1}{2} \varepsilon_{mnp} \varepsilon_{rs}
  \dual k^{n} F^{sp} + \frac{2}{3} H_{m} \dual h^{s} \mathcal{W}_{sr}
  +  \alpha \varepsilon_{mnp} \varepsilon_{rs} w^{tn} F^{sp} \dual
  h^{u} \mathcal{W}_{ut} \right]\ ,
\label{eq:fi-eom}\\
a_{m} &:& d \left[ \lambda \dual k^{m} - 2 \alpha  \lambda w^{rm}
\dual h^{s} \mathcal{W}_{sr} \right] +
\frac{4}{3} \lambda  F^{rm} \dual h^{s} \mathcal{W}_{sr} = 0\ , \label{eq:aeom} \\
b^{r} &:& d(\lambda {\cal W}_{rs}\dual h^{s}) = 0\ , \label{eq:beom} \\
\lambda &:& 1 + \Phi - \dual h^{r}\dual h^{s} \mathcal{W}_{rs} = 0\ ,
\label{eq:leom}
\end{eqnarray}
where we have left out the equations of motion for the transverse
scalars since these enter the duality discussion in a trivial manner.
We have also defined the quantities
\begin{eqnarray}
j_{rm} = \frac{\partial \Phi}{\partial w^{rm}}\ ,&& k^{m}
=\frac{\partial \Phi}{\partial f_{m}}\ .  \label{eq:defjk}
\end{eqnarray}

The last equation, (\ref{eq:leom}), is a constraint relating the
square of $h$ to the fields. It will be responsible for the
non-linearity of the duality relations.

The equations of motion for $b^r$, (\ref{eq:beom}), state that
$\lambda {\cal W}_{rs}\dual h^{s}$, which is also the electric field
conjugate to $b^r$, is constant. Assuming that it takes integer values
due to single-valuedness of wave functions \cite{Witten:1996im},
it can be interpreted
as a doublet of membrane charges $p_r$.

By inserting eq. (\ref{eq:beom}) into the equations of motion for the
$a$'s (\ref{eq:aeom}) and rewriting the LHS in terms of background
potentials we see that it is automatically satisfied if
\begin{eqnarray}
\label{eq:kis}
\dual k^{m} = 2 (\frac{2}{3} + \alpha) w^{rm} \dual h^{s}
\mathcal{W}_{rs}\ .
\end{eqnarray}
If this relation holds, these equations of motion are identical to the
Bianchi identities (\ref{eq:biwv1}) for the scalars $\phi$, \ie the
background fields enter in the same way on the right hand side.
This is the first of the implicit duality relations.
It can in turn be substituted into eq. (\ref{eq:fi-eom}) to yield
\begin{eqnarray}
&&d \left[ \lambda \dual j_{rm} - 2\alpha  \lambda f_{m}
\wedge \dual h^{s} \mathcal{W}_{sr} -  2 \beta  \lambda \varepsilon_{mnp}
\left( \varepsilon_{sr} \mathcal{W}^{vs} \mathcal{W}_{ut}
+ 2 \varepsilon_{st} \mathcal{W}^{vs} \mathcal{W}_{ur}  \right) w^{un} w^{tp}
\dual h^{s'} \mathcal{W}_{s'v} \right] \nn \\
&=& \lambda \left[ -\frac{2}{3} \varepsilon_{mnp}
\varepsilon_{rs} w^{un} F^{sp} \dual h^{t} \mathcal{W}_{tu} +
\frac{2}{3} H_{m} \dual h^{s} \mathcal{W}_{sr} \right] \ .
\end{eqnarray}
Demanding that this equation of motion in turn is automatically
satisfied, using the
Bianchi identities for $a$ (\ref{eq:biwv2}) and $\phi$
gives us the second duality relation
\begin{eqnarray}
\dual j_{rm} = 2\left( \alpha - \frac{1}{3}\right) f_{m} \dual h^{s}
\mathcal{W}_{sr} + \varepsilon_{mnp} \left( \left[ -2 \beta
+ \frac{1}{3} \right] \varepsilon_{rs} \dual h^{t'}
\mathcal{W}_{t't} w^{sn} w^{tp} \right. \nn \\
\left. + 2 \beta \varepsilon_{st} \dual h^{s} w^{un} w^{tp}
\mathcal{W}_{ur} \right)\ .
\label{eq:jis}
\end{eqnarray}
Note that the right hand sides of the equations of motion indeed are
integrable, which allows the identification of the implicitly defined
``conjugate variables'' $j$ and $k$ with the above expressions in
terms of field strengths.

Thinking of $\Phi$ as some non-linear expression whose lowest terms in
a power expansion in fields are proportional to $w^2$ and $f^2$, and
examining the content of eqs. (\ref{eq:kis}) and (\ref{eq:jis}), we note
that eq. (\ref{eq:kis}) is a non-linear duality relation between
$f$ and a certain projection of $w$, namely the one that ``points in
the same direction as $\star h^r$'' (in complex notation
$\hbox{Re}({\star\bar h}w)$). Eq. (\ref{eq:jis}), on the other hand,
contains two components, of which one, its contraction with $\star
h^r$, again is a non-linear duality
relation, while the other one, its contraction with
$\star h^t\epsilon_{tu}\mathcal{W}^{ur}$,
does not contain $f$ on the right hand side (these two components
correspond to the real and imaginary parts of ${\star\bar h}{\star j}$).
The counting of degrees of freedom tells us that the nine components of
$a_m$ and
$\phi^{rm}$ together represent only three physical degrees of
freedom. Therefore, the constraints imposed on their field strengths
should effectively contain two triplets of vectors, and the necessary
consistency condition (on $\Phi$) is that this indeed happens.

At this point, we have not been able to solve the system in complete
generality. Compared to situations encountered earlier, where the
equations analogous to (\ref{eq:kis}) and (\ref{eq:jis}) have been
simply a pair of duality relations, the system at hand is more
complicated. The non-linearity does not reside entirely in the
factor $\star h$, but appears also on the right hand side of
eq. (\ref{eq:jis}). The equations are very similar to those for $(p,q)$
 5-branes in type $IIB$, commented on in ref. \cite{Cederwall:1998ab}
and partially solved in a series expansion in
ref. \cite{Westerberg:1999fe}. The method of the latter of these
references is useful since it indicates that a unique solution exists.
The situation we have been able to handle exactly in the present case
is the one where the two duality relations are equivalent and the remaining
constraints are independent of these. This happens when
\begin{eqnarray}
\star h^r\epsilon_{rs}w^s=0\ ,
\label{eq:restriction}
\end{eqnarray}
\ie when all components of the vector
triplet $w$ point in the same direction as $h$. This is not a property
of the solution in the most
general situation, however, as it turns out to put some restriction on
the background fields, as will be seen below.

We will examine this restricted solution by starting from the assumption
that $h$ and $w$ are aligned, with respect to their $SL(2)$ indices. We
then note that if $\beta={1\over6}$, the $w^2$-terms on the right hand
side of eq. (\ref{eq:jis}) vanish. We get the relations
\begin{eqnarray}
\dual k^{m} &=& 2(\alpha+{2\over3})\dual h^{s} \mathcal{W}_{rs}
w^{rm}\ ,\nn \\
\dual j_{rm} &=& 2(\alpha-{1\over3})\dual h^{s} \mathcal{W}_{rs}
f_{m}\ .
\end{eqnarray}
The following calculations are simplified by the realisation that the
$SL(2)$ indices drop out completely when all fields point in one
direction.
We can then treat $h$ and $w$ as if they were single-component.
Define the two vector triplets $v$ and $u$ by
\begin{eqnarray}
w^{rm} &=& \frac{v^{m} \dual h^{r}}{\sqrt{\dual h^{s} \dual h^{t}
\mathcal{W}_{st} } }\ , \label{eq:ugly} \\
f_{m} &=& - \dual u_{m}
\end{eqnarray}
(in complex language, $w$ carries the same phase as $\star h$ and $v$
is its modulus: ${\star h}=|{\star} h|e^{i\chi}$, $w^m=v^me^{i\chi}$).
The duality relations between $f$ and $w$ now turn into algebraic
relations between $u$ and $v$ expressed as
\begin{eqnarray}
\frac{\partial\Phi}{\partial v} &=& 2(\alpha+{2\over3})\sqrt{1 +
  \Phi}\, u\ ,
\nn\\
\frac{\partial\Phi}{\partial u} &=& 2(\alpha-{1\over3})\sqrt{1 +
  \Phi}\, v\ ,
\label{eq:uvdual}
\end{eqnarray}
where the equation of motion for the Lagrange multiplier $\lambda$ has
been used in order to replace $h$ by the positive square-root.
The essential consistency check is that these two equations must
contain exactly the same information. This is a condition on the
function $\Phi$.
It is easily checked (for example by the first terms in a series
expansion) that the parameter $\alpha$ must take the value
$-{1\over6}$. Then the numerical factors on the right hand side of
eq. (\ref{eq:uvdual}) become 1 and $-1$. It is possible to simplify
the equations further: by the substitution $\rho=\sqrt{1+\Phi}$ they
turn into
\begin{eqnarray}
\frac{\partial\rho}{\partial v} &=& u\ ,
\nn\\
\frac{\partial\rho}{\partial u} &=&  -v\ .
\label{eq:uvdualtwo}
\end{eqnarray}
One has to remember, however, that $\Phi$ should not contain any
constant (independent of $u$ and $v$) term, which rules out trivial
solutions to eq. (\ref{eq:uvdualtwo}) 
like $\rho={1\over2}(v^2-u^2)$, $v=u$. In fact, it is $\Phi$, not
$\rho$, that turns out to be polynomial.

The problem of finding the ``right'' $\Phi$ can now be formulated as
follows:
We wish to obtain a $\Phi$ which when inserted into the two duality
relations above equates them. 
We have not been able to prove strictly that this requirement fixes
$\Phi$ uniquely, although a general series expansion in powers of $u$
and $v$, using an implementation of the methods below in Mathematica, 
indicates that this is
the case. 
Instead of pursuing that kind of general analysis, we will make an
Ansatz
for the duality relation directly by comparison with the ordinary
M2-brane wrapped on $T^{3}$.  Let us therefore turn to the dynamics
of the membrane to see how such a duality arises.

The internal scalars of the $M2$-brane compactified on $T^{3}$
enter the action according to
\begin{eqnarray}
S = \int d^{3} \xi \sqrt{-\det G}\ ,
\end{eqnarray}
with
\begin{eqnarray}
\det G_{\alpha\beta} = \det (g_{\alpha\beta} + v^{i}_\alpha
v_{i\beta})\ .
\end{eqnarray}
The field strength $v$ of the internal scalars is identical to the
pullback of the
internal vielbein $\hat e$ of eq. (\ref{eq:defofehat}).
In what will follow we do not write out the indices on $v$, viewing it
instead as a $3 \times 3$-matrix. This means not caring about the
signature of the world-volume metric; it turns out to be irrelevant
for these algebraic considerations. If we define the
invariants\footnote{We use a shorthand notation, where we, as
  mentioned, suppress the internal metric, and in addition omit
  transposition of matrices, so that, {\it e.g.}, $\tr v^2$ means
  $\tr(v^tv)$. The Cayley--Hamilton equation is most easily derived in
a frame where $v$ is diagonal.}
\begin{eqnarray}
T_{2} &=& \tr v^{2}\ , \\
T_{4} &=& \frac{1}{2} (\tr v^{4} - (\tr v^{2})^{2})\ , \\
T_{6} &=& \frac{1}{3} \tr v^{6} - \frac{1}{2} \tr v^{2} \tr v^{4}
+ \frac{1}{6} (\tr v^{2})^{3} = - (\det v)^{2}\ ,
\end{eqnarray}
we see that the $3\times 3$ matrix $v$ obeys the Cayley--Hamilton identity
\begin{eqnarray}
v^{6} = T_{6} + T_{4} v^{2} + T_{2} v^{4}\ ,
\end{eqnarray}
which implies that $\{ g,v,v^{2},v^{3},v^{4},v^{5} \}$ is a suitable
basis to express our duality in. The duality, now in a form that
relates vector to vector, is then given by
\begin{eqnarray}
 u = \frac{\partial \mathcal{L}}{\partial v} = -\sqrt{-\det G} (G^{-1}
 v)\ , \label{eq:memduality}
 \end{eqnarray}
which we will demand to be consistent with an action of the form
(\ref{eq:actanz}).
Therefore we make an Ansatz for $\Phi$ which is an arbitrary
polynomial of degree six in $v$ and $u$ (implying that the variation
of our Ansatz yields all the independent terms). Such a polynomial
will have $27$ independent coefficients which are determined by
demanding equivalence between the duality relation
(\ref{eq:memduality}) and each of the two relations (\ref{eq:uvdual}).
It is a priori not at all obvious that a $\Phi$ exists that fulfills
these relations. Seen as a series expansion in $u$ and $v$, they
contain an infinite number of equations for a finite number of
constants. Therefore it is very gratifying to verify that a solution
exists. It was obtained by implementing the Cayley--Hamilton relation
in Mathematica, and
is of the form\footnote{One linear combination of the constants
in the Ansatz remains undetermined. However, the function it
multiplies vanishes identically in $\Phi$ and its variations when
the duality relation holds. We have chosen the simplest version of $\Phi$.}
\begin{eqnarray}
\Phi&=&\frac{1}{2} \Bigl[\tr v^2 - \tr u^2
  +\frac{1}{2}(\tr(uv))^2-\tr(u^2v^2)\nn\\
&&\phantom{\frac{1}{2}\Bigl[}-\frac{1}{3}\tr v^6
+\frac{1}{2}\tr v^4\tr v^2-\frac{1}{6}(\tr v^2)^3
  \Bigr]\ .
\label{eq:Phiis}
\end{eqnarray}

It is worth mentioning
that in the case we have solved, both equations
(\ref{eq:kis}) and (\ref{eq:jis}) (given $\Phi$)
are non-linear equations involving
both variables, while eq. (\ref{eq:memduality})
represents a solution of $u$ in terms of
$v$. One may try to solve for $v$ in terms of $u$, but this turns out
to amount to solving a fifth order equation. Even with a solution
$v(u)$ at hand, one should not try to use $u$ (\ie $f$) alone to
describe membrane dynamics, since $f$ obeys a modified Bianchi
identity involving the scalars.

In the restricted solution presented above we have assumed that $w^r$ and
$h^r$ point in the same direction. The action is thus not covariant
under the  full $SL(2)$ group. Since we effectively have
only one $w$ and one $h$ out of the doublets, the actual symmetry is,
in a suitable basis, generated by one of the two generators of
$SL(2,{\mathbb Z})$, acting as ``$\tau\rightarrow\tau+1$''. The reason
for this is that the original input in our solution, namely
$\star h^r\epsilon_{rs}w^s=0$, only is a valid solution in certain
backgrounds. Acting on this equation (multiplied by $\lambda$) with an
exterior derivative yields the condition
\begin{eqnarray}
p_r\varepsilon^{rs}\mathcal{W}_{st}F^t=0\ 
\label{eq:restr2}
\end{eqnarray}
(where $p_r$ are the charges that arise when solving the equations of
 motion for $b_r$, as discussed in the
 earlier in this section), 
 putting restrictions on
the background. In addition, this restriction depends on which charge
sector we are considering. For such a background, equation
(\ref{eq:restriction}) and the calculation following it
constitute a valid solution to the relations (\ref{eq:kis}) and (\ref{eq:jis}).
However, due to the $p$-dependence of (\ref{eq:restr2}) it is not
meaningful to call this a U-duality invariant
formulation in such a background.
In order to claim U-duality covariance, we would need to restrict to a
smaller class of backgrounds contained in (\ref{eq:restr2}), namely
$F=0$. In such a background the action (\ref{eq:actanz}) with $\Phi$ given by
(\ref{eq:Phiis})
gives a U-duality covariant formulation of membrane
dynamics\footnote{The function $\Phi$ is given in terms of
  $v$, not $w$. Although the definition of $v$ in
  eq. (\ref{eq:ugly}) involves $\star h$ explicitly through it phase,
  this dependence cancels when each pair of $v$'s is replaced by a $w$
and a $\bar w$, or equivalently, by $\mathcal{W}_{rs}w^rw^s$.
This is of course necessary in order for the equation
of motion for the 2-forms to be unaffected.}.

In more general backgrounds, we would need to go back to the equations
(\ref{eq:kis}) and (\ref{eq:jis}) and make a more general Ansatz for
$\Phi$. It is quite clear from the structure of the non-linear terms
we earlier chose to discard from (\ref{eq:jis}) that $\Phi$ will then
contain odd as well as even powers of fields. A series expansion and
implementation of the Cayley--Hamilton identities using
\eg\ Mathematica could probably give the correct result, but we have
so far not been able to solve the equations.

An alternative way to finding the membrane dynamics would be to
consider $\kappa$-symmetry. One then considers the supermembrane
action given by the same formal expression, but with the background
fields being
pullbacks of superspace tensor fields.
For an unknown $\Phi$, the $\kappa$-variation is undetermined, but
just as for the equation of motion, assuming a duality relation turns
them into explicit expressions in terms of world-volume fields.
One then uses the
background dimension 0 and 1/2 values of the background tensor fields,
makes an Ansatz for the half-rank projector on
$\kappa$ (these tend to be especially simple in the present
formalism). The $\kappa$-invariance together with consistency of the
projection would then give the same duality relation as derived
above. This is not surprising, since the non-linear duality and
generalised chirality implied by $\kappa$-symmetry are intimately
linked together. We have not performed this calculation, but are
convinced that it will yield the same information, as was the case in
refs. \cite{Cederwall:1997ts,Cederwall:1998ab,Cederwall:1998gg}.

\section{Elimination of the 2-forms}

In order to eliminate the 2-forms and reformulate the dynamics
while retaining U-duality covariance,
we first note that the equation of motion (\ref{eq:beom})
for $b^{r}$ implies
\begin{equation}
\lambda \dual h^{s}\mathcal{W}_{rs} =p_r= {\rm constant}\ .
\end{equation}
This expression defines the two constants $p_{1}=p$ and
$p_{2}=q$, which in complete analogy with the string case
\cite{Witten:1996im,Townsend:1997kr,Cederwall:1997ts}
have to be integers in order for the quantum mechanical
wave-function to be single valued \cite{Witten:1996im}.
Inserting, for example, these $p_{r}$ into our implicit duality
relations then results in
\begin{eqnarray}
\lambda \dual k^{m} = w^{sm}p_{s}\ , \\
\lambda \dual j_{rm} = -f_{m} p_{r}\ ,
\end{eqnarray}
when restricted to the simpler set of backgrounds with $F=0$.
Note that for constant $(p,q)$ we have thus effectively broken the
$SL(2,\mathbb{Z})$-invariance,
although in an $SL(2,\mathbb{Z})$-covariant looking manner. This
means that we have restricted our action, which previously described
the entire $SL(2,\mathbb{Z})$-orbit of $(p,q)$-membranes, to describe
only one such membrane. Next we wish to derive the action for such a
membrane.

Define the field strength
\begin{eqnarray}
\tilde{w}^{m} \equiv p_{r} w^{rm} = \sqrt{p^2}v^m\ ,
\end{eqnarray}
of a field $\tilde{\phi}^m= p_{r}\phi^{rm}$ where the second equality follows from
(\ref{eq:ugly}).
Here,
$\sqrt{p^{2}}=\sqrt{p_{r}p_{s}\mathcal{W}^{rs}}=e^{\varphi/2}|p-q\tau|$,
and it is important for the derivation of the action and for the
interpretation of $\tilde\phi$ as an internal coordinate that we use a
linear combination of the $w^{rm}$ with {\it constant} coefficients that at
the same time is proportional to $v^m$.
Using the duality relation (\ref{eq:memduality}), the pair of Bianchi identities
(\ref{eq:biwv1}) and (\ref{eq:biwv2}) can be
turned into a pair of equation of
motion and Bianchi identity for the scalar $\tilde\phi$. Indeed
this is the definition of our duality. So by inserting
(\ref{eq:memduality}) into (\ref{eq:biwv2}) (using the relations
between the field strengths given above), the Bianchi identity for
$f$, we can eliminate $a$ and obtain the equation of motion for $\tilde{\phi}$
\begin{eqnarray}
d \Big[\sqrt{-\det G} \dual{}\Big(G^{-1}
\frac{\tilde{w}_m}{\sqrt{p^{2}}}\Big)\Big]
=-H_{m}
\ .
\end{eqnarray}
At the same time, the expression for the metric $G$ in terms of
$\tilde w$ is
\begin{eqnarray}\label{pqG}
G=g+{\tilde w\tilde w^t\over p^2}\ ,
\end{eqnarray}
so integrating
the equation of motion gives a
$(p,q)$-membrane action with a non-trivial scalar dependence in the tension:
\begin{eqnarray}\label{pqaction}
S_{(p,q)} =  -\int d^{3} \xi \sqrt{p^{2}}\sqrt{-\det G} +
\int \left((C^{r}-{1\over3}A^{rm}\wedge B_m)p_{r}+\tilde{w}^{m}
\wedge B_{m} \right)
\ ,
\end{eqnarray}
Here we have used the relation
$\delta \det M = \det M \tr(M^{-1} \delta M)$
for the first term of the action and added a $(p,q)$-covariant
3-form, following from the equations of motion for the coordinates,
to the WZ-term.

The kinetic term in the action tells us that the $(p,q)$ membrane tension, in
the Einstein frame, is proportional to $\sqrt{p^2}$. We would now like
to compare this kinetic term to the one in eq. (3.13)
of ref. \cite{Pioline:2004xq},
namely
\begin{equation}\label{sphervectSone}
S_1={\sqrt{\det[ZZ^t+(y^2+x_0^2){\mathbb I}]}
\over y^2+x_0^2}=\sqrt{y^2+x_0^2}\sqrt{\det[{\mathbb I}+\frac{ZZ^t}{y^2+x_0^2}]}\ .
\end{equation}
Here, $Z$ is the winding matrix, which after fixing world-volume
reparametrisations become identical to $\tilde w$, while $g$ on the
euclidean brane becomes equal to the identity matrix ${\mathbb
  I}$. Identifying $y^2+x_0^2$ with $p^2$ (at $\tau=i$)
means that the kinetic terms
are equal. The variables $x_0$ and $y$ is the pair of membrane charges
$p_r$. The corresponding actions also agree at arbitrary values of
$\tau$ (a more general form is also given in ref. \cite{Pioline:2004xq}).
The ``action'' of ref. \cite{Pioline:2004xq} also contains a
Wess--Zumino term (denoted $S_2$), which upon compactification of a $D=11$
membrane is obtained as the WZ term of a
membrane completely winding $T^3$ and thus coupling linearly to the
axion (a membrane instanton). In ref. \cite{Pioline:2004xq} it arises
from algebraic considerations, by demanding invariance under the larger
group $E_{6}$. We have not considered these aspects here. We comment on this
in the Conclusions.

We end this section with a direct check of the dilaton dependence of
the $(p,q)$ membrane tension.
According to eq. (\ref{pqaction}), the tension
for a (1,0)-membrane
divided by the tension for a (0,1)-membrane, for $a=0$, is given
by ${\rm T}_{(1,0)}/{\rm T}_{(0,1)}=g$.

Next, using results from section 2 we can give another argument
why ${\rm T}_{(1,0)}/{\rm T}_{(0,1)}=g$. We start by using that in
11 dimensions the Newton constant is defined as
\begin{equation}
2\kappa^{2}_{11}=(2\pi)^{8}\ell_{11}^{9}\ ,
\end{equation}
where $\ell_{11}$ is the 11-dimensional Planck length. In 8
dimensions we instead have defined the following Newton constant:
\begin{equation}
2\kappa^{2}_{8}=(2\pi)^{5}\ell_{8}^{6}\ ,
\end{equation}
where $\ell_{8}$ is the 8-dimensional Planck length. These
constants are related as follows:
\begin{equation}\label{118kappa}
2\kappa^{2}_{11}={\rm vol}(T^{3})2\kappa^{2}_{8}\ .
\end{equation}

Moreover, in the reduction above we used that $\sqrt{{\rm
det}G_{mn}}=e^{-\varphi}$. Hence, this implies that
\begin{equation}\label{vol}
{\rm vol}(T^{3})=(2\pi)^{3}g^{-1}\ell_{11}^{3}\ ,
\end{equation}
where $g$ is the closed string coupling constant. This means that
using (\ref{118kappa}) and (\ref{vol}), we obtain the following
relation between the Planck length in 11 dimensions and the Planck
length in 8 dimensions:
\begin{equation}\label{118Planck}
\ell_{11}=g^{-1/6}\ell_{8}\ .
\end{equation}

Next, from the membrane in 11 dimensions we obtain a
(1,0)-membrane with the same tension in units of the
11-dimensional Planck length. However, expressed in units of the
8-dimensional Planck length we instead obtain, using
(\ref{118Planck})
\begin{equation}\label{10M2}
{\rm
T}_{(1,0)}=\frac{1}{(2\pi)^{2}\ell_{11}^{3}}
=\frac{1}{(2\pi)^{2}g^{-1/2}\ell_{8}^{3}}\
.
\end{equation}
Furthermore, from the M5-brane we obtain, by wrapping it on the
3-torus, a (0,1)-membrane in 8 dimensions. The tension is
given by
\begin{equation}\label{01M2}
{\rm T}_{(0,1)}={\rm T}_{{\rm M}5}\times {\rm
vol}(T^{3})=\frac{{\rm
vol}(T^{3})}{(2\pi)^{5}\ell_{11}^{6}}=\frac{1}{(2\pi)^{2}g^{1/2}\ell_{8}^{3}}\
.
\end{equation}
Hence, using (\ref{10M2}) and (\ref{01M2}) we get that ${\rm
T}_{(1,0)}/{\rm T}_{(0,1)}=g$, which is the same as we obtained
above.

\section{Conclusions}
In this paper we have derived a membrane world-volume action
in 8 dimensions. 
The main result, explained in detail in section 3, 
is that the action is given by 
\begin{eqnarray}
S = \int d^{3} \xi \sqrt{-g} \lambda \left[1 + \Phi(w,f) -
  {\dual h^{r}}{\dual h^{s}} {\cal W}_{rs}\right]\ , \label{eq:actanz2}
\end{eqnarray}
where $w$, $f$ and $h$ are field strengths for scalars, vectors and
2-forms on the world-volume, including coupling to supergravity
background potentials, and where $\Phi$, for the restricted class of
background with vanishing 2-form field strengths, is given by eq. 
(\ref{eq:Phiis}). The coupling to background fields is consistent with
the duality to be imposed on the world-volume fields in addition to
the equations of motion encoded by the action.

8 is the highest dimensionality where the U-duality
group is ``non-trivial'', due to the emergence of an $SL(2)$ factor
relating metric and tensorial degrees of freedom. This is therefore
a suitable arena for trying to understand the origin of the spherical
vectors appearing in the work on membrane partition functions by
Pioline {\it et al.}
\cite{Pioline:2001jn,Pioline:2004xq},
in particular the role played by the extra degrees of freedom
on the membrane that appear to be required by their construction.

The method used here produces a theory that couples the
membrane to all fields in the supergravity background in a manifestly
U-duality symmetric way. The important property of this method is that
it associates an $n$-form field strength on the world-volume to every
$n$-form gauge potential in the background. Thus, in 8 dimensions the world
sheet theory will automatically contain two 2-form potentials
whose field equations can be solved producing two integration parameters
which can be identified as the charges appearing in the partition functions
of Pioline {\it et al.} \cite{Pioline:2001jn,Pioline:2004xq}.
This construction can in principle be generalized
to lower dimensions and bigger U-duality groups, although we expect
that the problems we had with solving the equations for general
backgrounds will persist or get worse.

This conclusion is reached by
comparing  our action for the $(p,q)$ membrane
and the classical action (see for example equation (3.13) in the
recent paper \cite{Pioline:2004xq} or (\ref{sphervectSone}) above)
 for the automorphic membrane.
This points strongly towards interpreting the ``extra'' dimensions of the
configuration space as the charges of the covariant membrane.
The $SL(2)$-charges in our action enter in the exactly same way as
the ``extra variables'' in the classical automorphic membrane action
at least for the part of the spherical vector of the automorphic membrane
that comes out of our analysis, namely the kinetic term $S_1$.

This lends some support for the conjecture in
\cite{Pioline:2001jn,Pioline:2004xq} that the
extra variables in the minimal representation based  on $E_6$ are really
related to membrane charges coming from 2-form potentials on the world-volume.
In our particular case the $(p,q)$ membranes originate from both
M2 and M5 branes
in 11 dimensions which follows from the fact that they
couple to 2-forms coming from both the 3-form and the 6-form potential in
11 dimensions as explained in detail above.
Note that the completely winding configurations
counted in the work of Pioline {\it et al.} are membrane instantons
which do not
arise from M5 branes winding the compact dimensions, at least not in the
standard way. This is related to the fact that the second part,
the Wess--Zumino type term $S_2$ in the spherical vector is not reproduced
in our work.
Further evidence for this interpretation can be found by repeating the argument
in lower dimensions.
In $D=7$, with U-duality group $SL(5,\RR)$, membrane charges come in
the fundamental representation, and can, from an 11-dimensional
perspective, be thought of as a membrane together with four types
of M5-branes, winding the four $T^3$ cycles in the $T^4$. The counting
agrees with the five ``extra'' variables taking part in the non-linear
realisation of $E_7$, thus lending further support to the conjecture in
 \cite{Pioline:2001jn,Pioline:2004xq}.

Although the observations made here makes the properties of membranes
a bit less mysterious from an algebraic point of view,
the understanding of the membrane as a ``fundamental''
microscopic building block of M-theory remains unclear. The analysis
made in this paper is purely classical. It does not show that the
proposal of ref. \cite{Pioline:2004xq} is correct, but it rather
elucidates what it amounts to, namely a sum over $(p,q)$ membrane
charge sectors.

\bibliographystyle{JHEP}
\bibliography{8dmem}

\end{document}